\documentclass{article}
\usepackage{spconf}
\usepackage{amsmath}
\usepackage{amssymb}
\usepackage{graphicx}
\usepackage{booktabs}
\usepackage[capitalise]{cleveref}
\usepackage{multirow}

\title{Uncertainty as a Predictor: Leveraging Self-Supervised Learning for Zero-Shot MOS Prediction}
\name{Aditya Ravuri\(^{1}\)\sthanks{This work was conducted during the author's internship at National Institute of Informatics, Japan.}, Erica Cooper\(^{2}\), Junichi Yamagishi\(^{2}\)}

\address{\(^{1}\) University of Cambridge, Cambridge, UK  \(^{2}\) National Institute of Informatics, Tokyo, Japan}
\begin{document}
%
\maketitle
\begin{abstract}
Predicting audio quality in voice synthesis and conversion systems is a critical yet challenging task, especially when traditional methods like Mean Opinion Scores (MOS) are cumbersome to collect at scale. This paper addresses the gap in efficient audio quality prediction, especially in low-resource settings where extensive MOS data from large-scale listening tests may be unavailable. We demonstrate that uncertainty measures derived from out-of-the-box pretrained self-supervised learning (SSL) models, such as wav2vec, correlate with MOS scores. These findings are based on data from the 2022 and 2023 VoiceMOS challenges. We explore the extent of this correlation across different models and language contexts, revealing insights into how inherent uncertainties in SSL models can serve as effective proxies for audio quality assessment. In particular, we show that the contrastive wav2vec models are the most performant in all settings.
\end{abstract}
\section{Introduction}
\label{sec:intro}

Mean Opinion Score (MOS) is a pivotal metric in assessing audio quality, particularly audio generated by synthesis and voice conversion systems \cite{voicemos22}. However, the traditional approach of collecting MOS data based on listening tests is fraught with challenges, as it is labour-intensive and subject to a range of contextual variables and potential artefacts in measurement \cite{voicemos22}. This underscores the imperative for an automated, reliable MOS prediction methodology.

The VoiceMOS 2022 and 2023 challenges \cite{voicemos22, voicemos23} have provided a unique opportunity to explore this domain, and self-supervised learning (SSL) based models have been found to be particularly performant. These challenges encompassed a diverse range of data, with each competition track providing natural and synthesized speech, along with MOS scores based on large-scale listening tests. The 2023 competition focused on zero-shot prediction of scores to emulate no/low-resource MOS score prediction; hence, no training data was provided. Many baseline models, for example, those mentioned in \cite{voicemos22}, use fine-tuned SSL models like wav2vec2 \cite{wav2vec2} with one additional layer for mapping the high dimensional latent to a MOS score.

This paper explores the surprising ability of \textbf{out-of-the-box} pre-trained SSL models on zero-shot MOS prediction. We are particularly inspired by approaches in biology where zero-shot prediction is possible using a model's uncertainty estimates, where uncertainties act as proxies for downstream tasks \cite{protein-zero-shot}. Our main hypotheses are that,
\begin{enumerate}
    \setlength{\parskip}{0cm} 
    \setlength{\itemsep}{0cm} 
    \item uncertainty estimates can be derived from the outputs of SSL models such as wav2vec, and that,
    \item these uncertainties can be used as proxies to MOS scores as high model uncertainty around the contents of an audio sequence must correspond to low audio quality.
\end{enumerate}
Our main contribution is to construct uncertainty measures, which we believe reflect an SSL model's overall audio-level uncertainty w.r.t. what (latent) tokens are uttered in an audio sample, and show that these measures correlate with audio quality. We also explore contexts within which these correlations are strongest. We also show, as other studies have shown \cite{wermos}, that measures of intelligibility correlate with MOS scores, and we show that our uncertainty measures correlate strongly with such intelligibility measures.

The observations in this paper are similar to those of \cite{ood-detection}, where it was noted that class probabilities of image classification models tend to be lower for out-of-domain images. Indeed, MOS score prediction can be seen as an out-of-domain audio classification task. We postulate that, in addition to learning acoustic, articulatory and semantic information, \cite{probe-1, probe-2}, SSL models learn a data distribution that helps evaluate quality (or out-of-domainness). Our work provides insight into why SSL models seem to work well for MOS prediction and provides a totally zero-shot baseline for low-resource MOS prediction settings.

\section{Audio-Level Uncertainty Measurement}
\label{sec:ums}

Given an audio sample, $\mathbf{a} \in \mathbb R^n$ of length $n$, we calculate a proxy for audio quality using a measure of uncertainty obtained from SSL models. Our reasoning is that audio samples that lead to significant uncertainties in the latent code/token probabilities output by SSL models such as wav2vec \cite{wav2vec} must be noisier and/or of poorer quality.

To obtain these measures of uncertainty, which we abbreviate UMs, we first pass the audio ($\mathbf{a}$) through an SSL model $v: \mathbb R^n \mapsto \mathbb R^{w, q}$, which outputs a $w, q$-sized matrix representation of the audio, where $w$ is the number of time windows and $q$ is the size of the latent vector. In the case of wav2vec, the outputs consist of contrastive predictive logits, and in the case of wav2vec2 based models with an ASR head (available in torchaudio), the outputs are logits corresponding to token utterance probabilities.

\begin{figure}[t]
    \centering
    \includegraphics[scale=0.6]{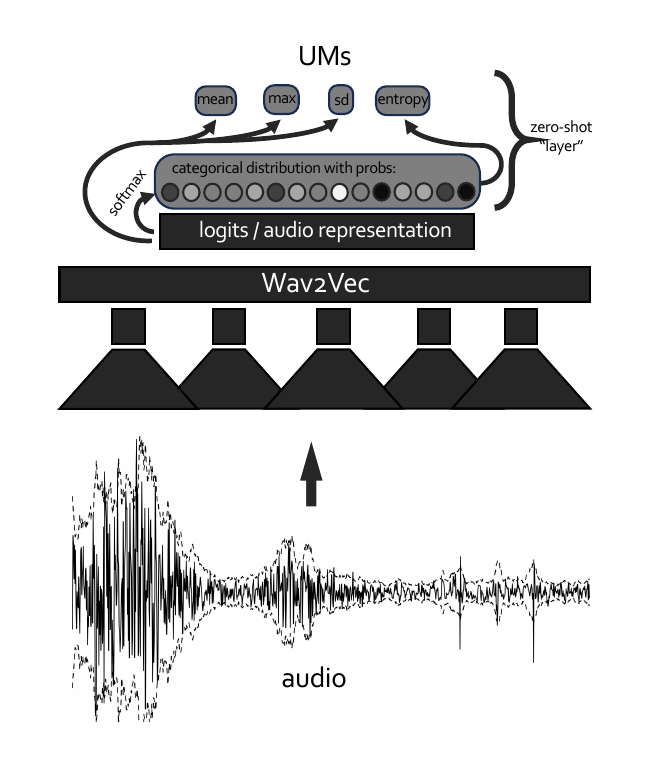}
    \vspace{-4mm}
    \caption{A diagrammatic representation of our zero-shot prediction pipeline, showing an audio sample fed to a wav2vec model, which emits log probabilities that can be used for evaluating the overall uncertainty of the model.}
    \label{fig:overview}
    \vspace{-3mm}
\end{figure}

In either case, as the softmax of these logits defines the probabilities of a categorical distribution, an entropy can be calculated. This corresponds to the amount of uncertainty an SSL model exhibits about which token is being uttered at any given time window (in the case of fine-tuned ASR models based on wav2vec2) or about which latent token is the most probable among a set of negatives (in the case where the outputs are wav2vec's contrastive predictive probabilities).

We calculate the entropy of the categorical over tokens/latent codes for each window and take an average over windows to obtain a measure of overall uncertainty for an audio file. That is,
\begin{align}
    &\text{UM[entropy]} (\mathbf{a})= \dfrac{1}{w} \sum_{i=1}^w H(L_i) \\
    &L_i \sim \text{Categorical}(p= \sigma(v(\mathbf{a})_i))  \;\; ^\forall i \in \{1, ..., w\}
\end{align}
where $H$ represents the entropy of a categorical random variable,
\begin{equation}
    H(L) = -\sum_x \mathbb P(L = x) * \log \mathbb P(L = x),
\end{equation}
and $\sigma$ represents the softmax operation.

In addition to entropy, we calculate the logits' mean, max and standard deviation as additional UMs, as we expect these to be computationally cheaper proxies to the entropy. The \textbf{maximum} logit is a good proxy as one would expect the entropy of a categorical distribution to be low if one of the logits is very high (which would represent high confidence in the emission of a token). Moreover, this measures the average confidence in the most likely token and is a comparable metric to that of \cite{ood-detection}. Hence, we expect the maximum and the entropy to be negatively correlated. Similarly, we would expect a high entropy if the \textbf{standard deviation} of the logits is low (as there is no preference for a token). The computation of these UMs is quite simple,
\begin{align}
    \text{UM[r]}(\mathbf{a}) &= \dfrac{1}{w} \sum_{i=1}^w r(v(\mathbf{a})_i),
\end{align}
where $r: \mathbb R^k \mapsto \mathbb R$ corresponds to a reduction such as a mean, standard deviation or maximum. Our prediction pipeline is illustrated in \cref{fig:overview}.

In the cases where a clear probabilistic interpretation of the output layer is not available, such as our experiments where a wav2vec2 model is available but not with a suitable ASR head or quantizer, we treat the encoder output as logits of a Categorical distribution, as these are typically fed to a quantizer.

Our UMs are similar to the metric of \cite{lmscore}, but we do not perform any additional training of the underlying models.

\section{Experimental Setup}
\label{sec:expt}

The VoiceMOS 2022 and 2023 challenges \cite{voicemos22, voicemos23} provide sets of audio files consisting of synthesised speech from text-to-speech systems and voice conversion systems, with each audio file being paired with a Mean Opinion Score (MOS) rating obtained through one of several listening tests, providing a benchmark for assessing audio quality. Metadata on the systems and listeners is available but we do not use this information in this preliminary work.

There are five main competition tracks across the two challenges that we consider, which we group as ``spoken English" and ``other" tasks. The spoken English track is the main track of the VoiceMOS 2022 challenge. The ``other'' tracks we consider are as follows,
\begin{itemize}
    \setlength{\parskip}{0cm} 
    \setlength{\itemsep}{0cm} 
    \item spoken Chinese tasks: these are the out-of-domain (OOD) track of VoiceMOS 2022 (which we abbreviate ``ood\_22'') and Track 3 from the VoiceMOS 2023 challenge (abbreviated ``ch\_23''),
    \item the spoken French task of VoiceMOS 2023 Track 1 (abbreviated ``fr\_23''),
    \item the English singing audio quality evaluation task of VoiceMOS 2023 Track 2, abbreviated ``sing\_23''.
\end{itemize}

For each task, we use various versions of wav2vec \cite{wav2vec} and wav2vec2 \cite{wav2vec2, xlsr} implemented in fairseq \cite{fairseq} and torchaudio \cite{torchaudio21, torchaudio23} as base SSL models, using which uncertainty measures (i.e. mean, max, standard deviation and entropy as described in \cref{sec:ums}) are computed for each audio sample. We then report the Spearman correlation (SRCC) between these measures and the associated MOS values for each task. We do not distinguish between train, test and validation splits and treat all datasets as test sets as our methodology concerns fully zero-shot prediction.

\section{Results and Discussion}
\label{sec:results}

We first ran our experimental setup on the spoken English data, using various base models. The results of this experiment are shown in \cref{tbl:eng-res}.

\begin{table}[tb]
\centering
\caption{Experimental results of our experimental setup using spoken English data}
\label{tbl:eng-res}
\begin{tabular}{@{}llrrcc@{}}
\toprule
model & type & mean & max & sd & entropy \\ 
\midrule
\multirow{2}{*}{wav2vec} & Large & \textbf{-69.1} & \textbf{68.2} & 64.0 & \textbf{-69.9} \\
& VQ & -60.1 & 58.4 & \textbf{68.3} & -69.0 \\
\midrule[\heavyrulewidth]
\multirow{5}{*}{\parbox{2cm}{wav2vec2 \\ ASR}} & base 10m & 6.5 & 23.4 & 26.8 & -23.0 \\
& base 100h & -23.8 & 45.5 & 46.3 & \textbf{-39.2} \\
& base 960h & \textbf{-45.7} & \textbf{50.6} & \textbf{52.4} & -37.3 \\
& large 960h & -9.9 & 39.9 & 40.3 & -25.7 \\
& lv60k 960h & 32.4 & -1.6 & 45.8 & -37.4 \\
& voxpop.\ en & -39.0 & 32.1 & 42.7 & -43.6 \\
\bottomrule
\end{tabular}
\end{table}

\subsubsection*{Uncertainty measures can be effective for zero-shot MOS prediction}

The wav2vec section of \cref{tbl:eng-res} shows that all our UMs correlate strongly with MOS scores, achieving SRCC values of about 70\%. The baseline model of \cite{voicemos22} achieved an SRCC of about 92\% on this task, so while our predictors are not state-of-the-art, they are nonetheless strong predictors given that no task-specific model training was done. The strong performance of the SSL models out of the box also may shed light on why these models always appear in performant solutions to MOS prediction.

The signs of SRCC between our UMs are consistent with our expectations (as high entropy would imply a low standard deviation and a high maximum) and are roughly equal in magnitude. There are some cases where there are differences between the performance of these UMs, however, and we leave this to future exploration.

\subsubsection*{UMs of wav2vec are stronger predictors of MOS than those of wav2vec2}

In comparing the performance of wav2vec and wav2vec2-based models, our experiments revealed a counter-intuitive trend, with UMs calculated using wav2vec models demonstrating better performance in MOS prediction.

We hypothesize that this may be due to two reasons. Firstly, wav2vec models generate contrastive predictive probabilities, which are specifically trained to identify a latent token from a set of negatives. High uncertainties implied by these probabilities are likely to directly reflect lower audio quality. In contrast, wav2vec2-based ASR models produce logits indicative of the probability of current token emissions. While we expect these to also have a correlation with audio quality, they might not be as directly or strongly related to audio quality as contrastive predictive probabilities. Secondly, we believe that wav2vec achieves better calibration for our task, in the sense that despite its lower capacity, UMs calculated using its outputs are more reflective of the model's uncertainties. Due to wav2vec2 being a better model, it may be unable to differentiate between audio samples with subtly differing quality. This second hypothesis is tested below.

\subsubsection*{Handicapping wav2vec2-based ASR models leads to a better performance}

Wav2vec2-based ASR models have three model components: a convolutional feature extractor, a transformer-based encoder and an ASR head. We devise a ``weaker'' form of these models by introducing a dropout layer between the feature encoder and the encoder of wav2vec, but crucially, inspired by (but unlike) the ideas of MC-dropout \cite{mcd}, we compute logits that are averaged across many forward passes of this network.

So, our new ``weaker'' wav2vec2-based ASR model passes each audio sample through its feature encoder, repeats this feature representation $k$ times (which we set at 100), drops out $p$ values randomly, and applies its encoder to each version of its $k$ feature representations affected by dropout. Finally, the $k$ output logits are averaged to obtain the final token logits.

\begin{figure}[t]
    \centering
    \includegraphics[scale=0.5]{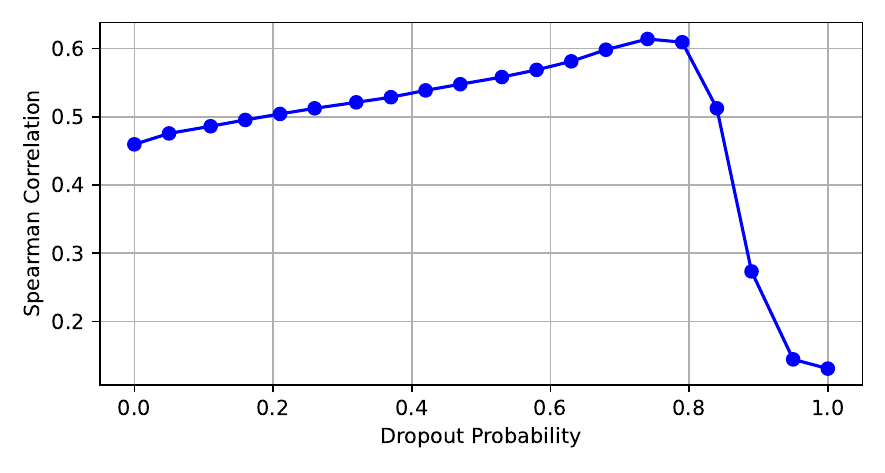}
    \vspace{-3mm}
    \caption{Plot of SRCC (measured against the spoken English MOS ratings) of our dropout-handicapped wav2vec2 Base 960H model against dropout probability, showing that performance \textbf{increases} initially as the model is handicapped.}
    \label{fig:dp}
    \vspace{-3mm}
\end{figure}

We observe that increasing the dropout probability from zero (equivalent to the original model) in our best-performing wav2vec2 model increases the SRCC by as much as 10\%; this is illustrated in \cref{fig:dp}.

\subsubsection*{wav2vec2 UMs are correlated with intelligibility measures}

For our best-performing model, i.e.\ wav2vec2 ASR Base 960H, we measured word error rates (WER) using transcripts generated by this model in conjunction with a 4-gram language model, which provides a measure of the intelligibility of audio. We found that our UMs have an average correlation of about 70\% with the WER and that these intelligibility measures have a correlation of about -53\% with MOS values.

\subsubsection*{Performance on the non-English datasets is poorer}

\begin{figure}[t]
    \centering
    \includegraphics[scale=0.55]{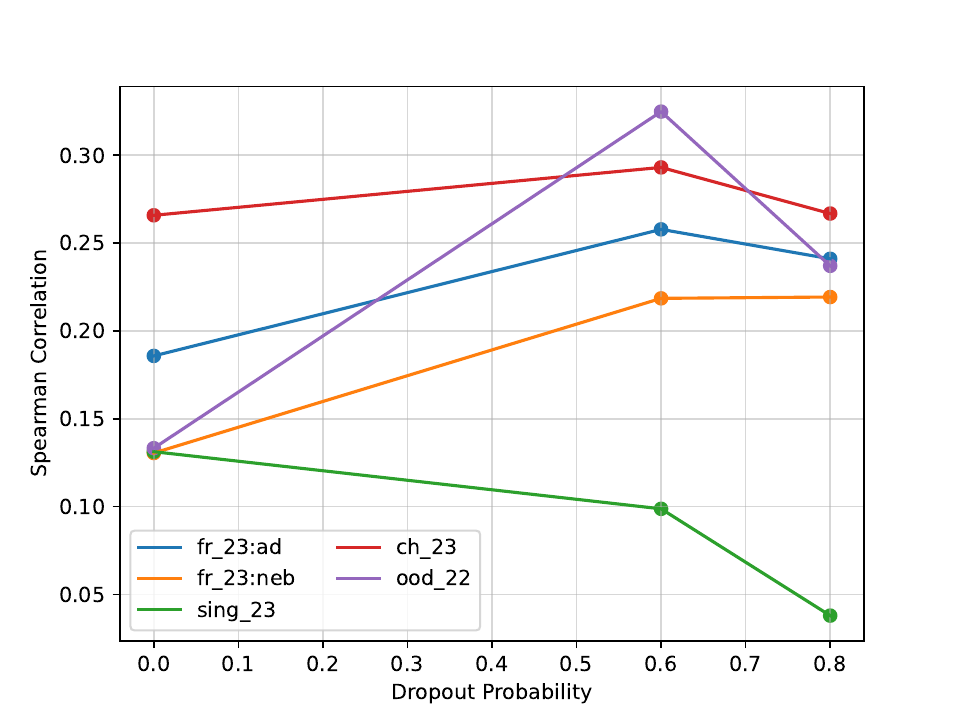}
    \vspace{-3mm}
    \caption{A plot of (average) SRCCs between our UMs and MOS ratings of ``other'' tasks, by (handicap) dropout probability. This figure shows that ``other'' task correlations are weaker than the spoken English case but they improve slightly with dropout.}
    \label{fig:ood}
    \vspace{-3mm}
\end{figure}

Fig.\ \ref{fig:ood} shows model performance on the following tasks by UMs of corresponding models,
\begin{itemize}
    \setlength{\parskip}{0cm} 
    \setlength{\itemsep}{0cm} 
    \item Spoken French (AD and NEB tasks, Track 1, VoiceMOS 2023) - we used the VOXPOPULI ASR BASE 10K FR SSL base model for this task.
    \item Spoken Chinese (OOD Track VoiceMOS 2022 and Track 3, VoiceMOS 2023) - we used the wav2vec2 XLSR53 SSL base model for this task.
    \item Sung English (Track 2, VoiceMOS 2023) - we used the wav2vec2 LARGE SSL base model for this task.
\end{itemize}
The performance of corresponding UMs in all cases is poor but weakly positive and is helped by model handicapping.

Interestingly though, as seen in \cref{tab:w2v-ood}, wav2vec (pre-trained on spoken English) are more performant than the zero-dropout versions of wav2vec2 models in each task.

\begin{table}[t]
\centering
\caption{Wav2vec performance on ``other'' tasks}
\label{tab:w2v-ood}
\begin{tabular}{@{}lcc@{}}
\toprule
Task & wav2vec\_large (\%) & vq-wav2vec.pt (\%) \\ 
\midrule
ood\_22 & 19.1 & 25.6 \\
fr\_23:ad & 25.8 & 20.7 \\
fr\_23:neb & 16.9 & 21.4 \\
sing\_23 & 19.3 & 27.9 \\
ch\_23 & 33.5 & 41.4 \\
\bottomrule
\end{tabular}
\end{table}

\subsubsection*{Pre-training wav2vec on target audio increases performance to $\sim$70\% SRCC}

Owing to the lack of out-of-the-box non-English wav2vec models, we further pre-train the existing English wav2vec-large checkpoint using high-quality target/evaluation data. This approach is practical as, often, one can obtain access to natural or almost natural audio samples that are comparable to the ones subject to evaluation. In each case, our training remains suitable for zero-shot prediction, as the model does not see any MOS scores associated with the audio. This pre-training increases wav2vec performance to above 70\% SRCC in the case of ood\_23, fr\_23 and the ch\_23 tasks, evaluated using the ``maximum UM''. The weakest performance is seen in the sing\_23 task, where the performance increases to about 50\% SRCC. Using all available evaluation data instead of just good-quality data for the ch\_23 task also increases performance to about 60\% SRCC, but for other datasets, performance may be dependent on the distribution of MOS scores within those datasets. This approach of (domain-adaptive) pre-training on target audio has also been noted to achieve better performance in \cite{dapt}.

\section{Conclusion}
\label{sec:conclusion}

This paper has demonstrated the potential of using uncertainty measures from SSL models for predicting MOS ratings, mirroring insight from other work in out-of-domain item classification. We find that wav2vec is notably the most performant for our tasks and we posit that this may be due to the nature of the contrastive predictive probability outputs and a better model calibration. Moreover, our fine-tuning experiment with wav2vec shows that a moderately strong predictor of audio quality can be achieved in no-resource settings. Future exploration could extend these insights to other domains where understanding out-of-domain data is critical, such as dimensionality reduction methods, which are known to have probabilistic interpretations reminiscent of self-supervised methods \cite{probdr}. In conclusion, our work contributes to the ongoing work in automated speech evaluation and understanding the working of SSL models for such tasks, showing applications of uncertainty quantification in this domain.

\vfill\pagebreak

\noindent
\textbf{Acknowledgments}\\
This study was supported by JST CREST Grant Number JPMJCR18A6 and MEXT KAKENHI grant 21K11951.
AR is supported by a studentship from the Accelerate Programme for Scientific Discovery. AR would also like to thank Simon Mathis for insightful discussions on out-of-domain identification tasks in biology.

\bibliographystyle{IEEEbib}
\bibliography{refs}

\begin{thebibliography}{10}

\bibitem{voicemos22}
Wen-Chin Huang, Erica Cooper, Yu~Tsao, Hsin-Min Wang, Tomoki Toda, and Junichi Yamagishi,
\newblock ``The voicemos challenge 2022,'' 2022.

\bibitem{voicemos23}
Erica Cooper, Wen-Chin Huang, Yu~Tsao, Hsin-Min Wang, Tomoki Toda, and Junichi Yamagishi,
\newblock ``The voicemos challenge 2023: Zero-shot subjective speech quality prediction for multiple domains,'' 2023.

\bibitem{wav2vec2}
Alexei Baevski, Henry Zhou, Abdelrahman Mohamed, and Michael Auli,
\newblock ``wav2vec 2.0: A framework for self-supervised learning of speech representations,'' 2020.

\bibitem{protein-zero-shot}
Joshua Meier, Roshan Rao, Robert Verkuil, Jason Liu, Tom Sercu, and Alexander Rives,
\newblock ``Language models enable zero-shot prediction of the effects of mutations on protein function,''
\newblock {\em bioRxiv}, 2021.

\bibitem{wermos}
Florian Hinterleitner, Steffen Zander, Klaus-Peter Engelbrecht, and Sebastian M{\"o}ller,
\newblock ``On the use of automatic speech recognizers for the quality and intelligibility prediction of synthetic speech,''
\newblock in {\em Konferenz Elektronische Sprachsignalverarbeitung}. TUDpress, Dresden, 2015, pp. 105--111.

\bibitem{ood-detection}
Dan Hendrycks and Kevin Gimpel,
\newblock ``A baseline for detecting misclassified and out-of-distribution examples in neural networks,'' 2018.

\bibitem{probe-1}
Hang Ji, Tanvina Patel, and Odette Scharenborg,
\newblock ``Predicting within and across language phoneme recognition performance of self-supervised learning speech pre-trained models,'' 2022.

\bibitem{probe-2}
Ankita Pasad, Ju-Chieh Chou, and Karen Livescu,
\newblock ``Layer-wise analysis of a self-supervised speech representation model,'' 2022.

\bibitem{wav2vec}
Steffen Schneider, Alexei Baevski, Ronan Collobert, and Michael Auli,
\newblock ``wav2vec: Unsupervised pre-training for speech recognition,'' 2019.

\bibitem{lmscore}
Soumi Maiti, Yifan Peng, Takaaki Saeki, and Shinji Watanabe,
\newblock ``Speechlmscore: Evaluating speech generation using speech language model,''
\newblock in {\em ICASSP 2023 - 2023 IEEE International Conference on Acoustics, Speech and Signal Processing (ICASSP)}, 2023, pp. 1--5.

\bibitem{xlsr}
Qiantong Xu, Alexei Baevski, and Michael Auli,
\newblock ``Simple and effective zero-shot cross-lingual phoneme recognition,'' 2021.

\bibitem{fairseq}
Myle Ott, Sergey Edunov, Alexei Baevski, Angela Fan, Sam Gross, Nathan Ng, David Grangier, and Michael Auli,
\newblock ``fairseq: A fast, extensible toolkit for sequence modeling,'' 2019.

\bibitem{torchaudio21}
Yao-Yuan Yang, Moto Hira, Zhaoheng Ni, Anjali Chourdia, Artyom Astafurov, Caroline Chen, Ching-Feng Yeh, Christian Puhrsch, David Pollack, Dmitriy Genzel, Donny Greenberg, Edward~Z. Yang, Jason Lian, Jay Mahadeokar, Jeff Hwang, Ji~Chen, Peter Goldsborough, Prabhat Roy, Sean Narenthiran, Shinji Watanabe, Soumith Chintala, Vincent Quenneville-Bélair, and Yangyang Shi,
\newblock ``Torchaudio: Building blocks for audio and speech processing,''
\newblock {\em arXiv preprint arXiv:2110.15018}, 2021.

\bibitem{torchaudio23}
Jeff Hwang, Moto Hira, Caroline Chen, Xiaohui Zhang, Zhaoheng Ni, Guangzhi Sun, Pingchuan Ma, Ruizhe Huang, Vineel Pratap, Yuekai Zhang, Anurag Kumar, Chin-Yun Yu, Chuang Zhu, Chunxi Liu, Jacob Kahn, Mirco Ravanelli, Peng Sun, Shinji Watanabe, Yangyang Shi, Yumeng Tao, Robin Scheibler, Samuele Cornell, Sean Kim, and Stavros Petridis,
\newblock ``Torchaudio 2.1: Advancing speech recognition, self-supervised learning, and audio processing components for pytorch,'' 2023.

\bibitem{mcd}
Yarin Gal and Zoubin Ghahramani,
\newblock ``Dropout as a bayesian approximation: Representing model uncertainty in deep learning,'' 2016.

\bibitem{dapt}
Wei-Cheng Tseng, Wei-Tsung Kao, and Hung yi~Lee,
\newblock ``Ddos: A mos prediction framework utilizing domain adaptive pre-training and distribution of opinion scores,'' 2022.

\bibitem{probdr}
Aditya Ravuri, Francisco Vargas, Vidhi Lalchand, and Neil~D. Lawrence,
\newblock ``Dimensionality reduction as probabilistic inference,'' 2023.

\end{thebibliography}

\end{document}